\documentclass[aps,amsmath,amssymb,showpacs,showkeys]{revtex4-2}
\usepackage[dvips]{graphicx}
\usepackage{times}
\usepackage{braket}
\usepackage{xcolor}
\usepackage{orcidlink}
\usepackage{hyperref}
\hypersetup{
  colorlinks=true,
  urlcolor=magenta,
  linkcolor=red,
  citecolor=blue
}
\begin{document}
\title{Analysing Ultra High Energy Cosmic Rays' Anisotropy in $\boldsymbol{f(R, T)}$ Gravity Theory}

\author{Swaraj Pratim Sarmah\orcidlink{0009-0003-2362-2080}}
\email[Email: ]{swarajpratimsarmah16@gmail.com}

\author{Umananda Dev Goswami\orcidlink{0000-0003-0012-7549}}
\email[Email: ]{umananda@dibru.ac.in}

\affiliation{Department of Physics, Dibrugarh University, Dibrugarh 786004, 
Assam, India}

\begin{abstract}
In this study, we investigated the anisotropy of diffusive Ultra-High Energy 
Cosmic Rays (UHECRs) by employing three cosmological models: two models from 
the $f(R, T)$ gravity theory and the other is the standard $\Lambda$CDM model. 
The primary objective of this work was to ascertain the role of the $f(R, T)$ 
gravity theory in comprehending the anisotropy of UHECRs without implicitly 
endorsing the conventional cosmology. We parameterized the magnetic field 
and the source distance in anisotropy calculations to align well with the 
observational data from the Pierre Auger Observatory for all the cosmological 
models. An uncertainty band is presented along with the $\chi^2$ test for all 
cosmological models to demonstrate the goodness of fitting. Our findings 
revealed that the amplitude of the anisotropy is highly sensitive to these 
cosmological models. Notably, the $f(R, T)$ models exhibited a lower amplitude 
of anisotropy (i.e., more isotropy), while the $\Lambda$CDM model predicted a 
comparatively higher amplitude at most of the energies considered.

\end{abstract}


\maketitle                                                                      

\section{Introduction}

It is generally believed that cosmic rays (CRs) with energies below 
$10^{17}$eV are mainly of galactic origin. Though their exact sources remain 
unknown, they are likely linked to sources such as supernova remnants or 
pulsars \cite{blasi201321, berezhko661, hewitt2015}. At ultra-high energies 
(UHEs), above $10^{18}$eV, CRs are thought to originate from extragalactic 
sources \cite{harari}. This idea is supported by several observations
\cite{mollerach2022, Auger2013, TA2017}. 
For instance, their arrival directions show no clear correlation with the 
distribution of galactic matter \cite{Auger2013, TA2017}. Another key finding 
is the detection of a dipolar pattern in the arrival directions of CRs with 
energies above $8$~EeV ($1$~EeV~$= 10^{18}$eV), which are pointing away from 
the galactic centre \cite{Auger2017a} further supporting their extragalactic 
origin. The energy spectrum and the arrival directions of CRs are crucial for 
understanding their nature. Changes in the slope of the spectrum can indicate 
shifts in propagation processes or transitions between different source 
populations, such as the dominance of extragalactic sources above the 
``ankle" around $5$~EeV, followed by a softening at approximately $13$~EeV 
\cite{AugerPRL2020, AugerPRD2020}, and a sharp decline starting near 
$50$~EeV \cite{Auger2021, icrc2021}. These shifts may also occur at 
intermediate energy levels. Furthermore, energy losses due to processes like 
pair production, as extragalactic protons interact with the cosmic microwave 
background (CMB) \cite{hillas677, blumenthal1596, berezinski_four_feat}, or 
due to diffusion effects \cite{lemoine_prd71}, can significantly shape the CR 
spectrum in the EeV energy range.

Significant progress has been made in recent years in the understanding of 
features of UHECRs. One major finding is a suppression in the CRs flux above 
$4 \times 10^{19}$~eV. This suppression has been confirmed by several 
experiments \cite{AugerPRL2020, hires2008, Auger2008, TA2013L1}. The decrease 
in flux at these energies is thought to result from the energy losses during 
propagation over cosmological distances, a phenomenon predicted nearly fifty 
years ago, known as the GZK cutoff \cite{KGreisen1966, GZ1966}. However, 
current data does not conclusively show whether the energy loss is the only 
factor causing this suppression. Additionally, researchers have set upper 
limits on the presence of photons \cite{Auger2009, TA2013, auger_jcap05}, 
neutrinos \cite{Auger2012L4, Icecube2013, auger_jcap10}, and neutrons 
\cite{Auger2012} in UHECRs. Data from the Pierre Auger Observatory (Auger) also 
indicates that at energies around EeV, CRs are primarily composed of light 
elements, mainly H and He, but their average mass progressively increases 
beyond a few EeV \cite{Auger2017, molerach, Auger2010, Auger2013JCAP, 
auger_prd9012, auger_prd90}.

As the origins of UHECRs remain unknown, studying the anisotropies in their
arrival directions may provide valuable clues in this regard 
\cite{mollerach201898, deligny_ptep2017, deligny_astrophys2019}. A significant 
challenge is that UHECRs, being charged particles, are deflected by magnetic 
fields in both galactic and extragalactic space. As a result, their arrival 
directions do not directly point to their sources. However, at the highest 
energies, these deflections become smaller due to the increased momentum of 
the particles, offering hope for identifying nearby, powerful extragalactic 
sources. Potential extragalactic sources include gamma-ray bursts, 
tidal disruption events, active galactic nuclei, and galaxy mergers 
\cite{mollerach2022}. The luminosity density of these sources, which depends 
on their distances (measured by redshift), affects the observed intensity of 
UHECRs and the production of secondary particles. Interactions with the CMB 
can cause energy loss, fragmentation of heavier particles, and production of 
photons and neutrinos through photo-pion interactions. These processes help to 
explain the UHECR spectrum at the highest energies and the anisotropies 
observed at intermediate angular scales. The large-scale anisotropies in UHECR 
arrival directions are linked to the uneven distribution of nearby galaxies 
within a few hundred Mpc \cite{Auger2018}. For energies above $4$~EeV, the 
dipolar anisotropy amplitude increases with energy. Auger has analysed over 
$2600$ UHECR events with energies above $32$~EeV, finding a deviation from 
isotropy at an intermediate angular scale with a $4\sigma$ confidence level
\cite{apj935}. Additionally, Auger data show a $4\sigma$ level of disfavoring 
of isotropy when correlated with starburst galaxies, highlighting their 
potential role in UHECRs anisotropy \cite{apjl853}. Similar findings have also 
been reported by the Telescope Array (TA) experiment 
\cite{TA_apjl898, TA_apj862, TA_arxiv14827, TA_auger_icrc}. Alongside the 
study of arrival direction patterns, understanding how UHECRs' composition 
changes with energy is equally important. Lighter and heavier components 
experience different deflection levels, providing valuable information. The 
Auger and TA are undergoing upgradation to improve their ability to investigate 
both anisotropies and composition of CRs at the highest energies.

One of the main challenges in modern cosmology is to explain the late-time 
cosmic acceleration \cite{reiss1998, perlmutter1999, spergel2003}. General 
relativity (GR), the most robust theory of gravity, struggles to naturally 
account for the accelerated expansion on cosmological scales, prompting 
exploration of alternative explanations. These alternatives primarily fall 
into two categories: dark energy (DE) and modifications or replacement of GR. 
GR links matter content with spacetime curvature, so any modification must 
alter one of these two components. DE introduces a form of matter with a 
repulsive gravitational effect \cite{brax2018}. Many theories of DE have been 
proposed, yet there is no experimental support for their validity. 
Almost parallelly, along with the DE the idea of the modification of the 
geometry part of GR has been developed to tackle the issue of cosmic expansion, leading to the modified theories of gravity (MTGs) \cite{sami, udg_prd}. As 
mentioned, since GR is a geometric theory of gravity, modifying its underlying 
geometry is a natural approach, forming the basis of modified gravity. 
A comprehensive review of these theories are provided in 
Ref.~\cite{nojiri2017}. A straightforward modification of GR replaces the 
Ricci scalar $R$ in the Einstein-Hilbert (EH) action with a function $f(R)$ 
of $R$, resulting in the $f(R)$ gravity theory. Some frameworks based on 
this approach were discussed in 
Refs.~\cite{felice2010,Sotiriou2010,udg_ijmpd,gogoi}. In passing it needs to
be mentioned that the gravity theories that intend to replace GR are based on
different geometries of spacetime than one based on GR. These theories
are usually termed as alternative theories of gravity (ATGs). One of the
recently most focused ATGs is the $f(Q)$ gravity theory \cite{jimenez, 
harko, sanjay1, sanjay2, noemi, psarmah}, which is based on the modified 
symmetric teleparallelism \cite{bahamonde_prd, bahamonde_plb}.  

Another promising approach of MTGs involves coupling geometry with matter. 
Non-minimal couplings (NMCs) \cite{azizi2014} address issues like 
post-inflationary preheating \cite{bertolami2011} and large-scale structure 
formation \cite{nesseris2009}. Scalar-tensor theories of gravity 
\cite{futamase1989, uzan1999} are based on the idea of MMC. Specifically, the 
$f(R, L_m)$ gravity \cite{harko2010} incorporates the matter Lagrangian $L_m$ 
into the action, while $f(R, T)$ gravity \cite{harko2011} generalises it 
further by including the trace $T$ of the energy-momentum tensor (EMT). The 
field equations in $f(R, T)$ gravity depend on a source term involving EMT 
variation, introducing coupling-induced acceleration and non-geodesic 
particle motion. Similar to some of the viable MTGs, the studies with 
$f(R, T)$ gravity span diverse areas like thermodynamics \cite{sharif2012}, 
scalar perturbations \cite{alvarenga2013} and dark matter implications 
\cite{zaregonbadi2016} of the Universe. Other studies with this gravity 
theory address the cosmic coincidence \cite{rudra2015} and gravitational 
waves \cite{sharif2019,jyotsna_dark}, showcasing its versatility in modeling 
cosmological phenomena.

Till now various research groups have employed diverse methods to investigate 
the anisotropy and propagation of CRs \cite{mollerach2022, harari, molerach, 
harari2021, merksch, m.ahlers, mollerech2022b, Abeysekara, grapes3, 
globus2019, G_sigl1999}. In our previous works, we have explored the flux 
characteristics \cite{swaraj1} and anisotropic properties \cite{swaraj2} of 
UHECRs for a single source within the framework of $f(R)$ gravity. 
Additionally, we have examined their propagation \cite{swaraj4} and flux 
suppression \cite{swaraj3} in this modified gravity theory along with the 
$f(Q)$ gravity theory for multiple sources. Building on these motivations, 
the current study aims to extend the investigation of CR anisotropies using an 
ensemble of sources, especially within the realm of $f(R, T)$ gravity, for 
the first time. For this purpose, we consider two models of $f(R, T)$ gravity, 
comparing their predictions with those of the standard $\Lambda$CDM model. 
To validate our results from the observational point of view, we utilize the 
surface detectors' data of Auger \cite{apj891}. The primary goal of this work 
is to assess the impact of $f(R, T)$ gravity on the anisotropic behaviour of 
UHECRs in comparison with predictions from standard cosmological models and 
constraining with observational data.

The rest of this paper is organized as follows. Section \ref{secII} is 
dedicated to the theoretical formalisms of the diffusion of UHECRs in 
turbulent magnetic fields (TMFs). Since this study considers the $f(R, T)$ 
gravity theory as the fundamental cosmological framework for investigating 
the anisotropy of UHECRs in galactic and extragalactic spaces, in Section 
\ref{secIII}, we present the required cosmological equations in this gravity 
theory. Section \ref{secIV} briefly discusses the $f(R, T)$ gravity models 
employed in this study. For an ensemble of sources, the CR anisotropy is 
analyzed in Section \ref{secV}. The numerical calculations of UHECRs 
anisotropies for the considered models, along with their comparison to 
observational data, are discussed in Section \ref{secVI}. Finally, we 
summarize our results and provide concluding remarks in Section \ref{secVII}.

\section{Turbulent Magnetic Fields and Diffusion of Cosmic Rays}\label{secII}
Due to some difficulties, it is hard to model the extragalactic magnetic 
fields \cite{han}. Their exact strengths are uncertain and they vary across 
different regions of extragalactic space \cite{hu_apj, urmilla}. At the 
centres of galaxy clusters, the field strength of these magnetic fields ranges 
from a few to tens of $\mu \text{G}$ \cite{han}. In less dense areas, they are 
weaker, typically between $1$ and $10$ nG, suggesting the presence of larger 
magnetic fields along cosmic structures like filaments. The coherence length 
$l_\text{c}$ is the maximum distance over which magnetic fields remain 
correlated. Galactic magnetic fields (GMF), with strengths of a few $\mu$G, 
have minimal impact on the CR spectrum due to their limited size, though 
they can influence the arrival directions of CRs. To simplify our study, we 
focus on the propagation of CRs in a turbulent and uniform extragalactic 
magnetic field. The key properties of this field are its root mean square 
(RMS) strength $B$ and coherence length $l_\text{c}$. Here, $B$ is defined as 
$\sqrt{\langle B^2(x)\rangle}$ and typically ranges from $1$ nG to $100$ nG 
\cite{feretti, Valle, Vazza}, while $l_\text{c}$ varies from $0.01$ Mpc to 
$1$ Mpc \cite{sigl}.

The effective Larmor radius of a charged particle with charge $Ze$ and energy 
$E$ in a magnetic field of strength $B$ is given by
\begin{equation}\label{larmor}
r_\text{L} = \frac{E}{ZeB} \simeq 1.1\, \frac{E/\text{EeV}}{ZB/\text{nG}}\;\text{Mpc}.
\end{equation}
A critical concept in charged particle diffusion is the critical energy 
$E_{\text{c}}$. It is defined as the energy at which the particle's Larmor 
radius equals the coherence length, i.e.~$r_\text{L}(E_{\text{c}}) = 
l_\text{c}$. The critical energy is expressed as
\begin{equation}\label{cri_energy}
E_\text{c} = ZeBl_\text{c} \simeq 0.9 Z\, \frac{B}{\text{nG}}\, \frac{l_\text{c}}{\text{Mpc}}\;\text{EeV}.
\end{equation}

In Ref.~\cite{harari}, using a numerical simulation the diffusion coefficient 
$D$ as a function of energy is provided as
\begin{equation}\label{diff_coeff}
D(E) \simeq \frac{c\,l_\text{c}}{3}\left[4 \left(\frac{E}{E_\text{c}} \right)^2 + a_\text{I} \left(\frac{E}{E_\text{c}} \right) + a_\text{L} \left(\frac{E}{E_\text{c}} \right)^{2-\gamma} \right],
\end{equation}
where $\gamma$ is the spectral index, and $a_\text{I}$ and $a_\text{L}$ are two 
coefficients. For a Kolmogorov spectrum \cite{molerach} in a TMF, $\gamma = 5/3$ with 
$a_\text{L} \approx 0.23$ and $a_\text{I} \approx 0.9$. The diffusion length 
$l_\text{D}$, representing the distance at which a particle's deflection is 
about one radian is defined as $l_\text{D} = 3D/c$. Using 
Eq.~\eqref{diff_coeff}, for $E/E_\text{c} \ll 0.1$, the diffusion length can 
be approximated as $l_\text{D} \simeq a_\text{L} l_\text{c} 
(E/E_\text{c})^{2-\delta}$ and for $E/E_\text{c} \gg 0.2$, it is $l_\text{D} 
\simeq 4 l_\text{c} (E/E_\text{c})^2$ \cite{molerach}.

In the diffusive regime, the transport equation for UHE particles is given as 
\cite{berezinkyGre}
\begin{equation}\label{diff_eqn}
\frac{\partial n}{\partial t} + 3 H(t)\, n - b(E,t)\,\frac{\partial n}{\partial E} - \frac{D(E,t)}{a^2(t)}\,\nabla^2 n = \frac{\mathcal{N}(E,t)}{a^3(t)}\,\delta^3({x}-{\bf{x}_\text{s}}),
\end{equation}
where $H(t) = \dot{a}(t)/a(t)$ is the Hubble parameter, $\dot{a}(t)$ is the 
time derivative of the scale factor $a(t)$, ${x}$ is the comoving coordinate 
and $r_\text{s}=({x}-{\bf{x}_\text{s}})$ is the source distance.
$n$ is the particle density and $\mathcal{N}(E)$ describes the 
number of particles emitted with energy $E$ per unit time. The energy losses 
(adiabatic and interaction losses), caused by cosmic expansion and 
interactions with the CMB are given by \cite{harari}
\begin{equation}
\frac{dE}{dt} = -\, b(E,t),\;\; b(E,t) = H(t)E + b_\text{int}(E).
\end{equation}
Here $b_{int}$ denotes the energy losses due to interaction with CMB and it 
includes both photopion and pair production. The general solution of 
Eq.~\eqref{diff_eqn} is \cite{berezinkyGre}
\begin{equation}\label{density}
n(E,r_\text{s})= \int_{0}^{z_{i}} dz\, \bigg| \frac{dt}{dz} \bigg|\, \mathcal{N}(E_\text{g},z)\, \frac{\textrm{exp}\left[-r_\text{s}^2/4 \lambda^2\right]}{(4\pi \lambda^2)^{3/2}}\, \frac{dE_\text{g}}{dE},
\end{equation}

where $E_\text{g}(E, z)$ represents the generation energy at 
redshift $z$ and $\mathcal{N}$ represents the source emissivity. 
The Syrovatskii variable $\lambda^2$ \cite{syrovatsy_1959} is defined as
\begin{equation}\label{syro}
\lambda^2(E, z) = \int_{0}^{z} dz \, \left| \frac{dt}{dz} \right| (1 + z)^2 D(E_g, z).
\end{equation}
Here $D(E_g, z)$ is the diffusion coefficient for the generation energy at $z$.
In the diffusive regime, the particle density increases with energy, distance 
from the source and TMF properties. This density enhancement describes how 
the CRs density evolves during propagation through intergalactic space and 
interactions with the CMB \cite{swaraj1}. It is quantified as the ratio of 
the density obtained from rectilinear propagation to actual density 
\cite{molerach}:
\begin{equation}\label{enhancement}
\xi(E,r_\text{s})=\frac{4\pi r_\text{s}^2 c\, n(E, r_\text{s})}{\mathcal{N}(E)}.
\end{equation}
For multiple sources, the source distance $r_\text{s}$ is replaced by 
$r_\text{i}$, which is discussed further in Section \ref{secV}.

\section{Basic equations of $\boldsymbol{f(R, T)}$ gravity}\label{secIII}
The total action of the $f(R,T)$ gravity is given by \cite{harko2011}
\begin{equation}\label{actionRT}
S = \frac{1}{2\kappa} \int f(R,T) \sqrt{-g} \, d^4x + \int \mathcal{L}_m \sqrt{-g} \, d^4x, 
\end{equation}  
where $\kappa = 8\pi$, $R$ is the Ricci scalar, $T$ is the trace of of the 
energy-momentum tensor $T_{\mu\nu}$, $g$ is the determinant of the metric 
tensor $g_{\mu\nu}$ and $\mathcal{L}_m$ is the matter Lagrangian. As discussed 
earlier, $f(R,T)$ is a function of both $R$ and $T$. Here we used the 
geometrized unit system wherein $G = c = 1$.
Varying the action \eqref{actionRT} with respect to the metric tensor 
$g_{\mu\nu}$ yields the field equations of $f(R,T)$ gravity as given 
by \cite{harko2011}
\begin{equation}\label{field}
f_R(R,T) R_{\mu\nu} - \frac{1}{2} f(R,T) g_{\mu\nu} + \left( g_{\mu\nu} \Box - \nabla_\mu \nabla_\nu \right) f_R(R,T) = \kappa T_{\mu\nu} 
- f_T(R,T) T_{\mu\nu} - f_T(R,T)\, \Theta_{\mu\nu},
\end{equation}  
where $f_R = \partial f/\partial R$, $f_T = \partial f/\partial T$, 
$\nabla_\mu$ is the covariant derivative and $\Box = \nabla^\mu \nabla_\mu$ 
is the d’Alembertian operator. The tensor $\Theta_{\mu\nu}$ depends on the 
matter Lagrangian and simplifies for a perfect fluid as \cite{harko2011}  
\begin{equation}\label{thetaPerfect}
\Theta_{\mu\nu} = -\,2T_{\mu\nu} + p g_{\mu\nu}.
\end{equation}  
and the energy-momentum tensor is defined as
\begin{equation}\label{energymomentum}
T_{\mu\nu} = -\,\frac{2}{\sqrt{-g}} \frac{\delta (\sqrt{-g} \mathcal{L}_m)}{\delta g^{\mu\nu}},
\end{equation}  
with its trace $T = g^{\mu\nu} T_{\mu\nu}$.
In a spatially flat Friedmann-Lemaître-Robertson-Walker (FLRW) spacetime with 
metric  
\begin{equation}\label{frwmetric}
ds^2 = -\,dt^2 + a^2(t) \left(dx^2 + dy^2 + dz^2\right),
\end{equation}  
the modified Friedmann equations are
\begin{align}\label{frw1}
3H^2 f_R + \frac{1}{2}\left(f - R f_R\right) + 3H \dot{f}_R & = \left(\kappa + f_T\right) \rho,\\[5pt]
\label{frw2}
2\dot{H} f_R + \ddot{f}_R - H \dot{f}_R & = -\left(\kappa + f_T\right) \rho,
\end{align}  
where $H = \dot{a}/a$ is the Hubble parameter as mentioned earlier, $\rho$ and
$p$ are the total energy density and pressure of the Universe.
These two equations can be expressed as effective Friedmann equations as
follows \cite{rudra2021_npb}:
\begin{align}
\label{fe1}
3H^2 & = \kappa\, \rho_\text{eff} = \kappa(\rho + \rho_\text{mod}),\\[5pt] 
\label{fe2}
2\dot{H} + 3H^2 & = -\,\kappa \left(\rho_\text{eff} + p_\text{eff}\right), 
\end{align}
where $\rho_\text{eff} = \rho + \rho_\text{mod}$ and 
$p_\text{eff} = p + p_\text{mod}$ (for pressure less dust $p=0$) with 
$\rho_\text{mod}$ and $p_\text{mod}$ are contributions to the energy density 
andpressure of the Universe from the modified gravity, effectively behaving 
like a dark fluid component as given by \cite{rudra2021_npb}
\begin{equation}
\rho_\text{mod}=\frac{-f(R,T)-6H\dot{f_{R}}+2\left(\kappa+f_{T}\right)\rho+f_{R}\left(R-2\kappa\rho\right)}{2\kappa
f_{R}}
\end{equation}
and
\begin{equation}
p_\text{eff}=p_\text{mod}=\frac{3\left[\ddot{f_{R}}+f(R,T)+5H\dot{f_{R}}-Rf_{R}-\left(f_{T}+\kappa\right)\rho\right]}{3\kappa f_{R}}.
\end{equation}  
Generally, the trace of the energy-momentum tensor for a perfect fluid is 
given by $T=\rho + 3p$. However, we can rewrite this expression for the 
pressure-less dust as $T=\sigma \rho$, where $\sigma$ is a parameter of 
adjustment.

\section{$\boldsymbol{f(R, T)}$ Gravity Models}\label{secIV}
This section outlines the $f(R, T)$ gravity cosmological models used 
to compute the cosmological parameters required for this study. In particular, 
we consider two models of $f(R, T)$ gravity that include both minimal and 
non-minimal coupling between curvature and matter. Also, the expression for 
the Hubble parameter $H(z)$ for these models is presented in this section.

\subsection{Model I: Minimal coupling in exponential form}
The first $f(R, T)$ gravity model we have considered here is given by 
\cite{rudra2021_npb}
\begin{equation}\label{mod2}
f(R,T)=\alpha R+\beta e^{T}\!,
\end{equation}
where $\alpha$ and $\beta$ are two model parameters. For this
model the first Friedmann Eq.~\eqref{fe1} takes the form \cite{rudra2021_npb}:
\begin{equation}\label{mod22}
H^{2}(z)=\frac{1}{6\alpha}\left[2\left(\rho_{m0}+\rho_\text{rad0}\left(1+z\right)\right)\left(1+z\right)^{3}+\beta\left(2\rho_\text{m0}\left(1+z\right)^{3}-1\right)e^{\sigma\rho_\text{m0}\left(1+z\right)^{3}}\right],
\end{equation}
where $\rho_\text{m0}$ and $\rho_\text{rad0}$ are matter and radiation 
components of the energy density $\rho$ of the Universe. The respective 
dimensionless density parameters are defined as
\begin{equation}\label{denpara}
\Omega_\text{m0}=\frac{\rho_\text{m0}}{3H_{0}^{2}},\;\;\;\Omega_\text{rad0}=\frac{\rho_\text{rad0}}{3H_{0}^{2}}.
\end{equation}
Using these dimensionless density parameters Eq.~\eqref{mod22} can be rewritten
as
\begin{equation}\label{mod222}
H(z)=\frac{H_{0}}{\sqrt{2\alpha}}\left[2\left(\Omega_\text{m0}+\Omega_\text{rad0}\left(1+z\right)\right)\left(1+z\right)^{3}+\beta\left(2\Omega_\text{m0}\left(1+z\right)^{3}-\left(3H_{0}^{2}\right)^{-1}\right)e^{3H_{0}^{2}\sigma
\Omega_\text{m0}\left(1+z\right)^{3}}\right]^{1/2}\!\!\!.
\end{equation}
In this work we used the values of the model parameters as $\alpha=0.8469$,
$\beta = -0.4285$, and $\sigma=-0.0041$. These parameter values represent
the best-fit results obtained using cosmic chronometer (CC) and Baryon 
Acoustic Oscillation (BAO) data \cite{rudra2021_npb}.

\subsection{Model II: Non-minimal coupling}
The second model we have considered is a model of non-minimal coupling between 
matter and gravity, and is given by \cite{rudra2021_npb}
\begin{equation}\label{mod4}
f(R,T)=R+f_{0}R\, T^{\delta}\!,
\end{equation}
where $f_{0}\neq 0$ and $\delta$ are parameters of this model. For this model, 
the first Friedmann Eq.~\eqref{fe1} can be expressed as \cite{rudra2021_npb}

\begin{align}
\nonumber
H^{2}(z)=\, & \frac{\left[\sigma\left(\rho_\text{m0}+\rho_\text{rad0}\left(1+z\right)\right)+f_{0}\left(\sigma\rho_\text{m0}\left(1+z\right)^{3}\right)^{\delta}\left(\sigma\rho_\text{m0}+\rho_\text{rad0}\left(\sigma-\delta\right)\left(1+z\right)\right)\right]}{\sigma+f_{0}\left(3\delta+\sigma\right)\left(\sigma\rho_\text{m0}\left(1+z\right)^{3}\right)^{\delta}}\, \times\\[5pt]
\label{mod44}
&\left(1+z\right)^{3} \left[\frac{4\sigma}{3\left\{\sigma+f_{0}\left(\sigma-\delta\right)\left(\sigma\rho_\text{m0}\left(1+z\right)^{3}\right)^{\delta}\right\}}-\frac{1}{1+f_{0}\left(\sigma\rho_\text{m0}\left(1+z\right)^{3}\right)^{\delta}}\right].
\end{align}
By using the dimensionless density parameters from Eq.~\eqref{denpara}, we 
can write the above equation in the form:
\begin{align}
\nonumber
H(z)=\,& H_{0}\frac{\left[3\sigma\left(\Omega_{m0}+\Omega_{rad0}\left(1+z\right)\right)+3f_{0}\left(H_{0}^{2}\right)^{\delta}\left(3\sigma\Omega_{m0}\left(1+z\right)^{3}\right)^{\delta}\left(\sigma\Omega_{m0}+\Omega_{rad0}\left(\sigma-\delta\right)\left(1+z\right)\right)\right]^{1/2}}{\left[\sigma+f_{0}\left(3\delta+\sigma\right)\left(3H_{0}^{2}\sigma\Omega_{m0}\left(1+z\right)^{3}\right)^{\delta}\right]^{1/2}}\, \times\\[5pt]
\label{mod444}
&\left(1+z\right)^{3/2}\left[\frac{4\sigma}{3\left\{\sigma+f_{0}\left(\sigma-\delta\right)\left(3H_{0}^{2}\sigma\Omega_{m0}\left(1+z\right)^{3}\right)^{\delta}\right\}}-\frac{1}{1+f_{0}\left(3H_{0}^{2}\sigma\Omega_{m0}\left(1+z\right)^{3}\right)^{\delta}}\right]^{1/2}\!\!\!\!.
\end{align}
The values of the model parameters we have used for this study are 
$f_{0}=2.4285$, $\sigma=0.1836$ and $\delta = 0.1002$ \cite{rudra2021_npb}. 
The cosmological parameters used in this work, based on WMAP 7 
\cite{WMAP} results for the $\Lambda$CDM model, are $H_0 = 72$ km s$^{-1}$ Mpc$^{-1}$, $\Omega_{\text{m}0} = 0.3$, and $\Omega_{\text{r}0} =  10^{-4}$. 
The cosmological time evolution with respect to redshift that appears in 
Eq.~\eqref{density} can be expressed as \cite{swaraj1}
\begin{equation}
\bigg | \frac{dt}{dz} \bigg | =\frac{1}{(1+z)\, H}.
\end{equation}
Using this equation with the expressions of the Hubble parameter for the model
considered above, the cosmological time evolution with respect to redshift for 
the corresponding cosmological models can be calculated. 
\begin{figure}[!h!]
\centerline{
\includegraphics[scale=0.5]{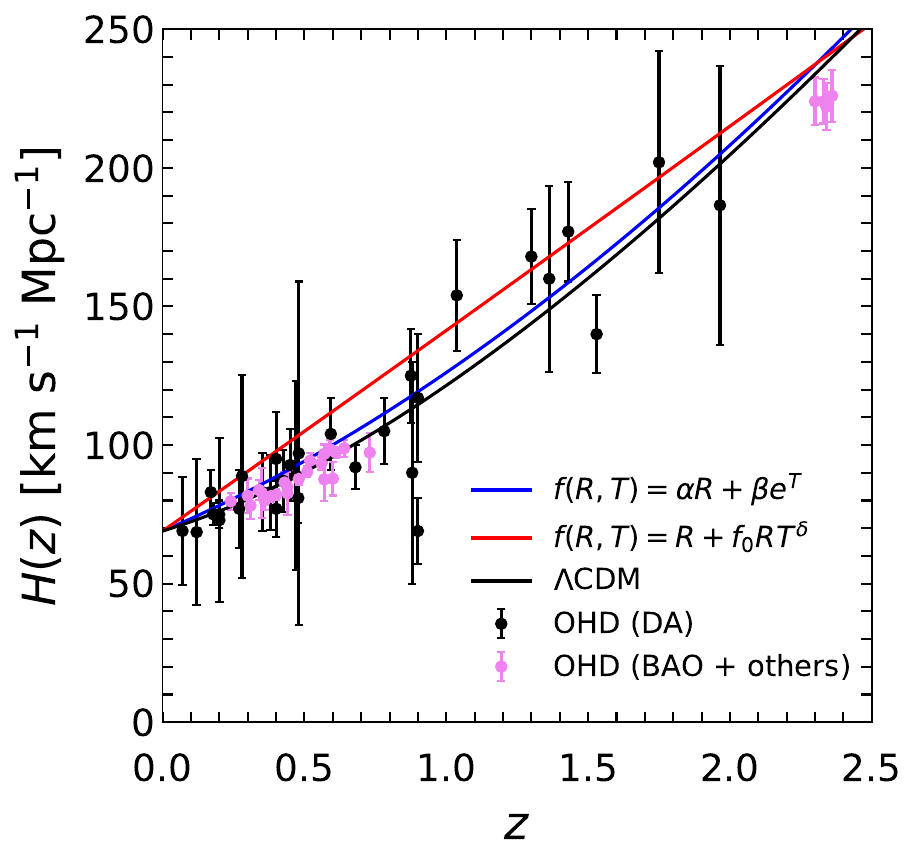}}
\vspace{-0.2cm}
\caption{Variations of Hubble parameter $H(z)$ with redshift $z$ as predicted 
by the $f(R, T)$ gravity models along with the $\Lambda$CDM model, and in 
comparison with the observational Hubble data (OHD) obtained from Differential 
Age (DA) and BAO methods 
\cite{solanki, swaraj1}.}
\label{hz}
\end{figure} 

We compare the Hubble parameter results obtained for both $f(R, T)$ gravity 
models, and the standard $\Lambda$CDM model along with the observational 
Hubble data (OHD) acquired from Differential Age (DA) and BAO methods \cite{solanki, swaraj1} in Fig.~\ref{hz}. 
The plot demonstrates that both $f(R, T)$ gravity models' results fit well 
with the observational data.

\section{Anisotropy of Cosmic Rays for an ensemble of sources}\label{secV}
The flux from a CR source at a distance $r_\text{s}$, much greater than the 
diffusion length $l_\text{D}$ can be calculated by solving the diffusion 
equation in the expanding Universe \cite{berezinkyGre}. The resulting 
expression is given as \cite{Manuel}
\begin{equation}\label{fluxeq}
J(E) = \frac{c}{4\pi} \int_{0}^{z_{\text{max}}} \!\! dz \, \left| \frac{dt}{dz} \right| \, \mathcal{N}\left[E_\text{g}(E, z), z\right] \frac{\exp\left[-r_\text{s}^2 / (4 \lambda^2)\right]}{(4 \pi \lambda^2)^{3/2}} \frac{dE_\text{g}}{dE},
\end{equation}
where $z_{\text{max}}$ is the maximum redshift at which the CRs emitting by
a source can be perceived.
Since we focus on multiple sources rather than a single source, we employ a 
relation given by \cite{moleachjcap,Manuel},
\begin{equation}
r_\text{i} = \left(\frac{3}{4\pi}\right)^{\!1/3}\!\!\! d_\text{s}\, \frac{\Gamma(i + 1/3)}{(i - 1)!},
\end{equation}
where $d_\text{s}$ represents the distance between the sources and $i$ 
indicates the $i$-th source from the average distance. The term $d_\text{s}$ is linked with the source density 
$n_\text{s}$ as $d_\text{s} \approx n_\text{s}^{-1/3}$. Consequently, for a 
discrete source distribution, summing over the sources results in a specific 
factor \cite{moleachjcap, Manuel}:
\begin{equation} \label{F_supp}
F \equiv \frac{1}{n_\text{s}} \sum_i \frac{\exp\left[-r_\text{i}^2/4 \lambda^2\right]}{(4\pi \lambda^2)^{3/2}}.
\end{equation}
Subsequently, in Eq.~\eqref{fluxeq}, after summing all the sources, we can 
express the modified flux for an ensemble of sources in terms of the Hubble 
radius $R_\text{H} = c/H_0$ for the first $f(R, T)$ gravity model as
\begin{align}
J(E) \Big |_\text{mod-1}\simeq  
&\, \frac{R_\text{H}\, n_\text{s}}{4\pi} \int_{0}^{z_\text{max}} \!\! dz\, (1+z)^{-1}
 \times \left[ \frac{1}{\sqrt{2\alpha}} \left\{ 2 \left( \Omega_{m0} + \Omega_{rad0} \left( 1+z \right) \right) \left( 1+z \right)^{3} \right. \right. \nonumber \\[5pt]
&\left. \left. +\, \beta \left( 2\Omega_{m0} \left( 1+z \right)^{3} - \left( 3H_{0}^{2} \right)^{-1} \right) 
e^{3H_{0}^{2}  \sigma \Omega_{m0} \left( 1+z \right)^{3}} \right\}^{1/2} \right] \mathcal{N} \left[ E_\text{g}(E, z), z \right] \, \frac{dE_\text{g}}{dE} \,F.
\end{align}

Similarly, for the second $f(R, T)$ gravity model, the modified flux can be 
expressed as follows:
\begin{align}\label{flux2}
J(E) \Big |_\text{mod-2} \simeq  
&\,\frac{R_\text{H}\, n_\text{s}}{4\pi} \int_{0}^{z_\text{max}} \!\! dz\, (1+z)^{-1}\, \times  \nonumber \\[5pt] 
& \left[ \frac{\left\{ 3\sigma \left( \Omega_{m0} + \Omega_{rad0}(1+z) \right) 
+ 3f_{0}\left(H_{0}^{2}\right)^{\delta} \left(3\sigma\Omega_{m0}(1+z)^{3}\right)^{\delta}\left(\sigma\Omega_{m0} + \Omega_{rad0}(\sigma - \delta)(1+z)\right) \right\}^{1/2}} {\left[\sigma + f_{0}(3\delta+\sigma) \left(3H_{0}^{2}\sigma\Omega_{m0}(1+z)^{3}\right)^{\delta}\right]^{1/2}}
\right. \times \nonumber \\[5pt]
& \left. (1+z)^{3/2} \left[\frac{4\sigma}{3 \left\{\sigma + f_{0}(\sigma-\delta) \left(3H_{0}^{2}\sigma\Omega_{m0}(1+z)^{3}\right)^{\delta}\right\}} 
- \frac{1}{1 + f_{0} \left(3H_{0}^{2}\sigma\Omega_{m0}(1+z)^{3}\right)^{\delta}}
\right]^{1/2} 
\right] \times \nonumber \\[5pt]
& \mathcal{N}\left[E_\text{g}(E, z), z\right]\,\frac{dE_\text{g}}{dE}\,F.
\end{align}
Furthermore, we can rewrite Eq.~\eqref{syro} in terms of the Hubble radius 
$R_\text{H}$ and using Eq.~\eqref{diff_coeff} as
\begin{equation}\label{ad}
\lambda^2(E,z)= H_0\frac{R_\text{H} l_\text{c}}{3}\int_{0}^{z}\!\!dz\, \bigg | \frac{dt}{dz} \bigg |\,(1+z)^2 \left[4 \left(\frac{(1+z)\,E}{E_\text{c}} \right)^2 + a_\text{I} \left(\frac{(1+z)\,E}{E_\text{c}} \right) + a_\text{L} \left(\frac{(1+z)\,E}{E_\text{c}} \right)^{2-\gamma}   \right].
\end{equation}
Now we calculate the density enhancement factor of CR protons for the 
$f(R, T)$ gravity models. Using this result, we compute the CR protons' flux 
and ultimately their anisotropy as predicted by the two $f(R, T)$ gravity 
models. The anisotropy is evaluated following the methodology outlined in 
Ref.~\cite{Supanitsky} and which is expressed as  
\begin{equation} \label{aniso}
\Delta = 3 ~ \frac{\eta}{\xi},
\end{equation}
where $\eta$ is the modification factor as given by (see 
Refs.~\cite{swaraj1, swaraj2})
\begin{equation}
\eta = \frac{J(E)}{J_{0}(E)}.
\end{equation}
Here, $J_{0}(E)$ is the CR flux without any kind of energy losses and it is 
given by
\begin{equation}\label{flux0eq}
J_{0}(E) = \frac{c}{4\pi} \int_{0}^{z_{\text{max}}} \!\! dz \, \left| \frac{dt}{dz} \right| \, \mathcal{N}_{z\rightarrow 0}(E) \frac{\exp\left[-r_\text{s}^2 / (4 \lambda^2)\right]}{(4 \pi \lambda^2)^{3/2}}.
\end{equation} 
Utilizing these aforementioned relations within the framework of the 
contemplated models of MTGs, we will analyse the numerical outcomes pertaining 
to the anisotropy in the ensuing section.

\section{Results and discussion}\label{secVI}
This section focuses on numerical computations, data fitting, and analysis of
results. The \texttt{python scipy} and \texttt{numpy} libraries are used 
for the numerical calculations, while \texttt{matplotlib} is used for creating 
plots. All the plots presented assume the primary particles as protons with a 
spectral index $\gamma = 2$ and the redshift $z=2$.  All the parameters for the 
cosmological models have been adopted from Ref. \cite{rudra2021_npb}, as mentioned earlier.

\begin{figure}[h!]
\centerline{
\includegraphics[scale=0.5]{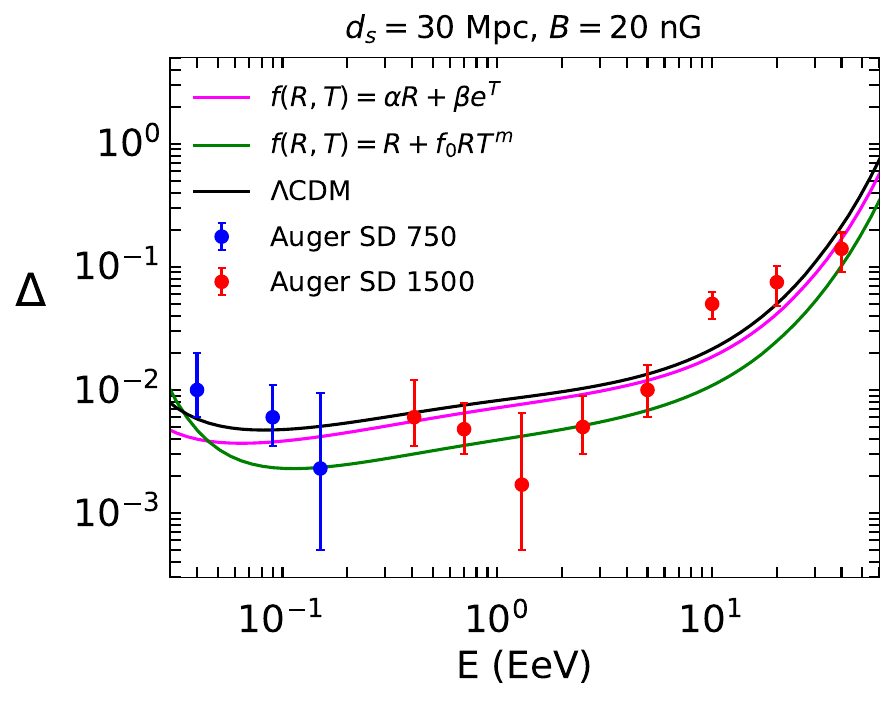}}
\vspace{-0.2cm}
\caption{The anisotropy of UHECRs as a function of energy $E$ for two 
$f(R, T)$ gravity models along with the standard $\Lambda$CDM model for 
$d_\text{s} =30$ Mpc and $B=20$ nG. The observational data are taken from the 
Auger \cite{apj891}.}
\label{aniso1}
\end{figure}

Fig.~\ref{aniso1} presents the UHECRs anisotropy parameter $\Delta$ as a 
function of energy $E$ (in EeV) for different theoretical models, compared 
with the Auger SD 750 (blue points) and SD 1500 (red points) datasets. The 
magenta and green lines represent the first and second $f(R, T)$ gravity models 
respectively, and the black line corresponds to the standard $\Lambda$CDM 
model. For this plot the values of the magnetic field ($B$) and source 
separation distance ($d_\text{s}$) are fixed at $20$ nG and $30$ Mpc 
respectively, for all cases. 
The anisotropy prediction of the $\Lambda$CDM model shows a steady increase 
from $\Delta \sim 4\times 10^{-3}$ at low energies ($<0.1$ EeV) to higher 
values beyond ${10}$ EeV. The experimental data align reasonably well with 
the curve at low energies. The $f(R, T) = \alpha R + \beta e^{T}$ 
model predicts a similar qualitative trend but shows a slight suppression of 
anisotropy at low energies compared to the $\Lambda$CDM. The 
$f(R, T) = R + f_0 R T^{\delta}$ model also exhibits anisotropy suppression 
compared to $\Lambda$CDM at intermediate ($>0.1$ EeV) and high energies 
($>10$ EeV). At low energies, all three models provide a similar trend of 
predictions, which align well with the experimental data. At intermediate and 
high energies, the $f(R, T)$ gravity models deviate from the $\Lambda$CDM 
prediction, potentially reflecting modified gravity effects. However, 
significant scatter in the high energy experimental data and deviations from 
theoretical trends suggest the need for further refinements in both 
observational and theoretical approaches. We will take the fitting parameters 
of the $\Lambda$CDM model as reference and based on them, modify those 
parameters for the $f(R, T)$ gravity model.

\begin{figure}[h!]
\centerline{
\includegraphics[scale=0.5]{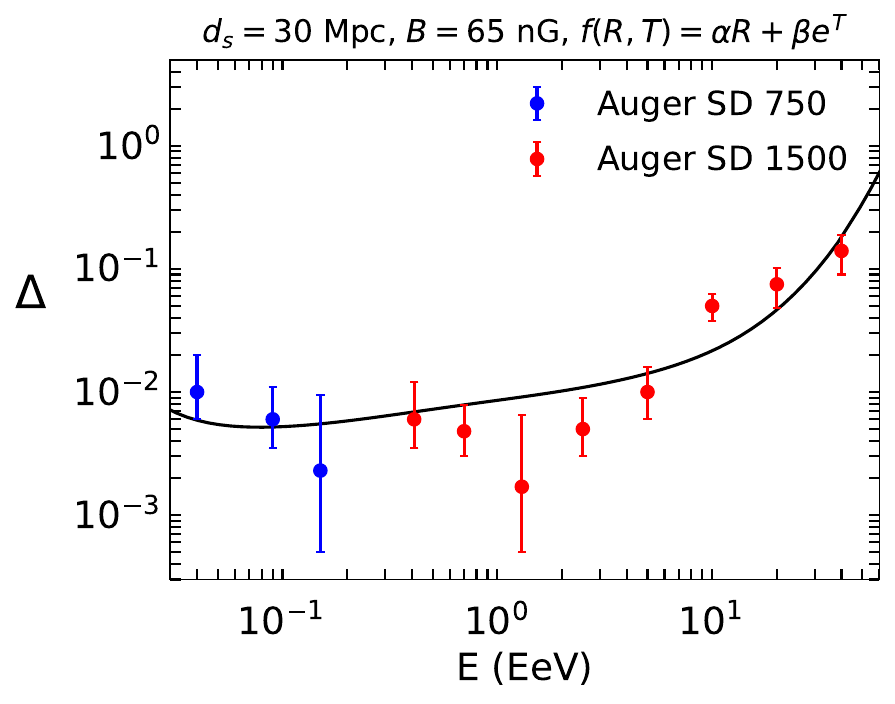}\hspace{0.5cm}
\includegraphics[scale=0.5]{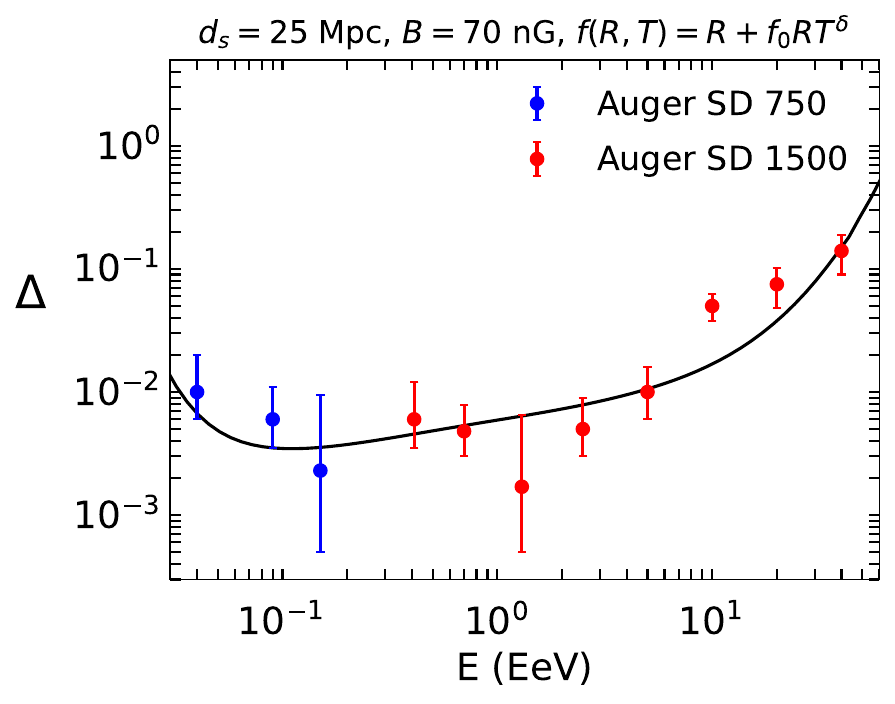}}
\vspace{-0.2cm}
\caption{Left: The modified anisotropy of UHECRs as a function of energy $E$ 
for the $f(R, T)= \alpha R + \beta e^{T}$ model with $d_\text{s} =30$ Mpc and 
$B=65$ nG. Right: The modified anisotropy of the UHECRs as a function of 
energy $E$ for the $f(R, T)=R + f_0 R T^{\delta}$ model with $d_\text{s} =25$ 
Mpc and $B=70$ nG. The observational data are taken from the 
Auger \cite{apj891}.}
\label{aniso2}
\end{figure}

Fig.~\ref{aniso2} shows the updated theoretical predictions for the two 
$f(R, T)$ gravity models, where adjustments in the magnetic field ($B$) and 
source separation distance ($d_\text{s}$) are performed to improve alignment 
with the Auger SD 750 and SD 1500 datasets. In the left panel, corresponding 
to the model $f(R, T) = \alpha R + \beta e^{T}$, the parameters are set to 
$d_\text{s} = 30$ Mpc and $B = 65$ nG. The theoretical curve maintains a similar 
qualitative trend to the previous figure but exhibits a slightly better match 
with the experimental data, particularly at low and intermediate energies. 
In the right panel, for the model $f(R, T) = R + f_0 R T^{\delta}$, the 
parameters are adjusted to $d_\text{s} = 25\ \mathrm{Mpc}$ and $B = 70\ \mathrm{nG}$. 
The resulting theoretical curve aligns more closely with the Auger SD 750 and 
SD 1500 data compared to the previous figure, especially at low energies. 
While in the high energy rise in anisotropy remains a common feature of both 
models, the adjustments lead to reduced discrepancies with the experimental 
data. The chi-squared ($\chi^2$) values for models' predictions are 
calculated using the formula:
\begin{equation}
\chi^2 = \sum \left( \frac{O_i - M_i}{\sigma_i} \right)^2\!\!,
\end{equation}
where $O_i$ is the $i$th observed value, $M_i$ is the corresponding 
model-predicted value, $\sigma_i$ is the uncertainty in the observation.
The $\chi^2$ and  the reduced $\chi^2$ values for the considered models
are shown in Table \ref{tab:chi2_values}.
\begin{table}[h!]
\centering
\caption{$\chi^2$ and $\chi_\text{red}^2$ values for the three models under consideration.}
\vspace{3pt}
\begin{tabular}{l@{\hspace{20pt}}c@{\hspace{20pt}}c}
\hline
\textbf{Model} & \textbf{ $\chi^2$} & \textbf{$\chi_\text{red}^2$} \\
\hline
Minimal coupling & 20.596 & 0.982 \\
Non-minimal coupling & 18.286 & 0.871 \\
$\Lambda$CDM & 16.552 & 0.788 \\
\hline
\end{tabular}
\label{tab:chi2_values}
\end{table}
\begin{figure}[h!]
\centerline{
\includegraphics[scale=0.5]{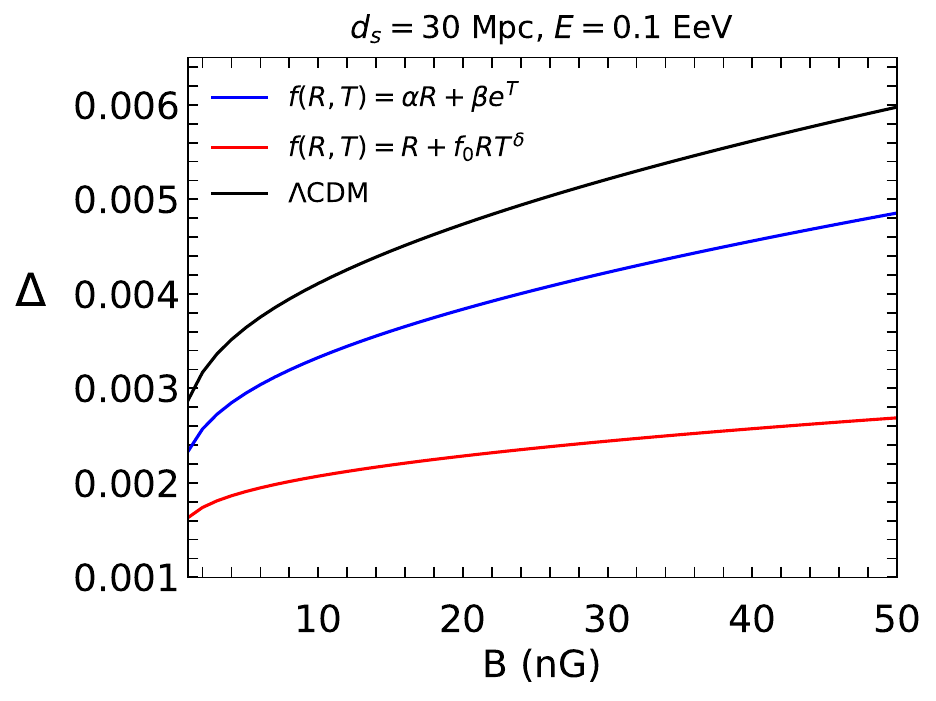}\hspace{0.5cm}
\includegraphics[scale=0.5]{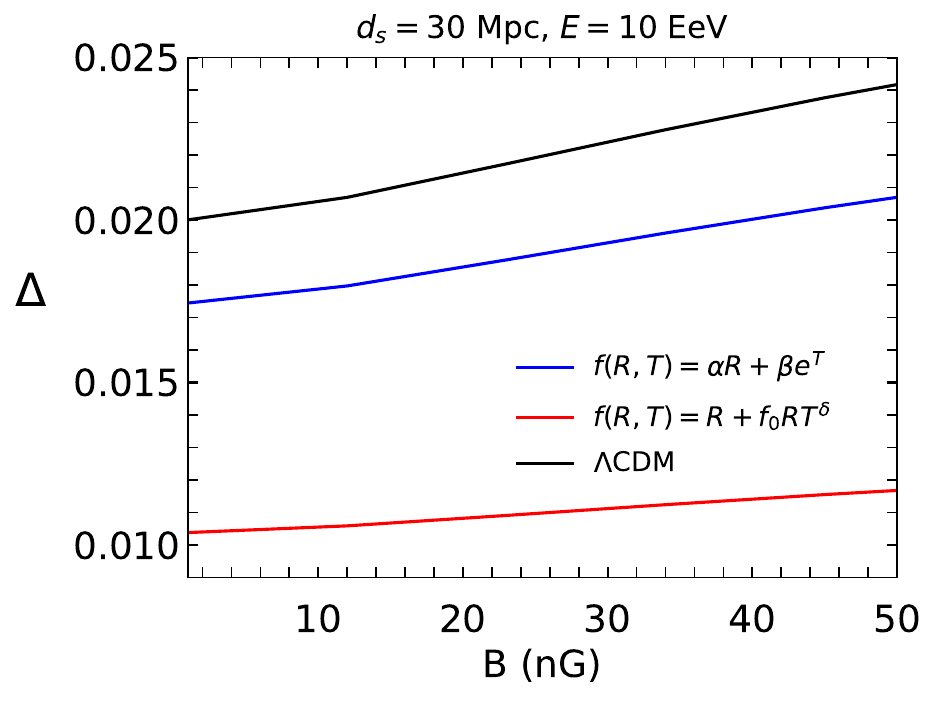}}
\vspace{0.3cm}
\centerline{
\includegraphics[scale=0.5]{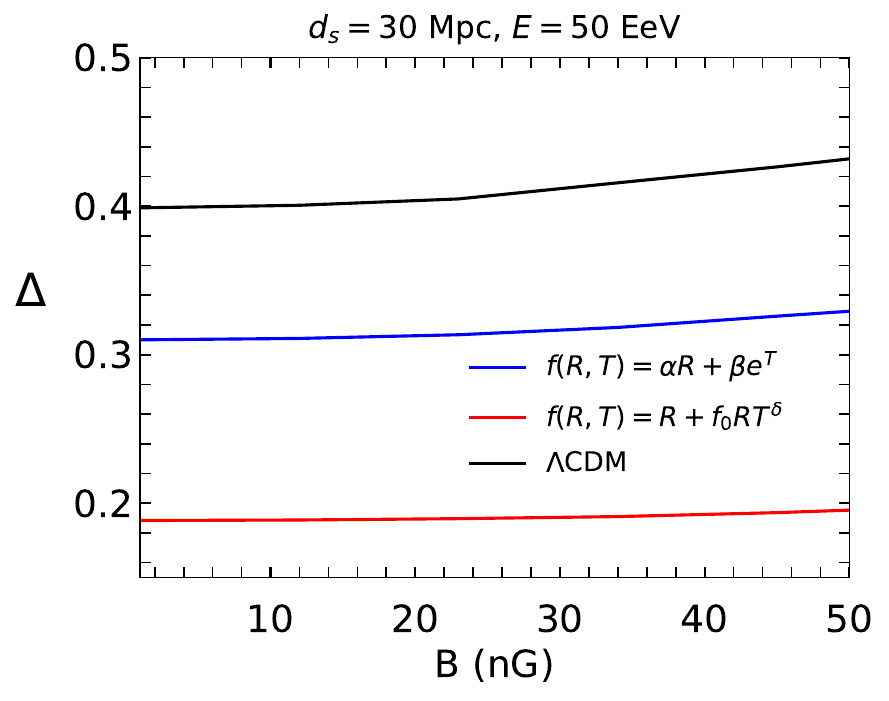}}
\vspace{-0.2cm}
\caption{The anisotropy of the UHECRs as a function of the magnetic field $B$ 
for two $f(R, T)$ gravity models along with the standard $\Lambda$CDM model 
with $d_\text{s} =30$ Mpc and $E=0.1$ EeV (left), $10$ EeV (right), and $50$ 
EeV (bottom).}
\label{aniso3}
\end{figure}

Fig.~\ref{aniso3} depicts the variation of the anisotropy parameter $\Delta$ 
as a function of the magnetic field strength $B$ (in nG) for three specific 
CR energies $E = 0.1$ EeV, $E = 10$ EeV, and $E = 50$ EeV in the left, right 
and bottom panels respectively. The three curves in each plot correspond to 
different cosmological models: $\Lambda$CDM (black),  
$f(R,T) = \alpha R + \beta e^{T}$ (blue) and 
$f(R,T) = R + f_0 RT^{\delta}$ (red). Across all models and energies, $\Delta
$ increases with $B$, reflecting the impact of stronger magnetic fields on CRs' 
deflection. For a fixed $B$, the anisotropy parameter $\Delta$ increases with 
energy, as higher energy CRs experience reduced deflection and retain 
more directional information as it shows a flattened pattern. Among the models, 
$\Lambda$CDM predicts the largest anisotropy for all energies, followed by the 
$f(R, T) = \alpha R + \beta e^{T}$ model, while the 
$f(R, T) = R + f_0 RT^{\delta}$ model exhibits the lowest values of $\Delta$. This 
indicates that the underlying cosmological model influences the predicted 
anisotropy, with MTG  models generally predicting smaller anisotropy compared 
to the standard $\Lambda$CDM framework.

These parameter adjustments highlight the sensitivity of anisotropy predictions to different factors such as the magnetic field and source separation distance. 
The improved alignment of the theoretical predictions with the data 
underscores the potential of modified gravity models to explain CRs' anisotropy
while demonstrating the need for precise tuning of model parameters.

\section{Conclusion}\label{secVII}
In this work, we have studied the CRs' anisotropy parameter $\Delta$ as a 
function of energy $E$ in the UHE range in the framework of two $f(R, T)$ 
gravity models in comparison to the standard $\Lambda$CDM model. To account 
for the effects of different parameters on the anisotropy, we have explored 
various magnetic field strengths ($B$) and source distance separations 
($d_\text{s}$). The comparison of the CRs' anisotropy parameter $\Delta$ across 
different theoretical models ($\Lambda$CDM and two $f(R, T)$ gravity models)
highlights the potential of modified gravity theories to affect in the CRs 
study. The adjustments to the magnetic field strength ($B$) and source 
separation distance ($d_\text{s}$) in the $f(R, T)$ models have demonstrated 
significant improvements in aligning theoretical predictions with the Auger 
SD 750 and SD 1500 datasets. 

The $\Lambda$CDM model provides a reasonable qualitative description of the 
anisotropy trends at both lower and higher energies with the observed data. In 
contrast, in the same range of magnetic field and source separation, the 
$f(R, T) = \alpha R + \beta e^{T}$ model captures the of anisotropy 
at higher energies more effectively but does not fit well in the low energies. 
The $f(R, T) = R + f_0 R T^{\delta}$ model further refines the alignment at low 
energies but depicts a deviation at high and intermediate energy with 
observational trends. Thus, the magnetic field and separation distance 
modification is required for the $f(R, T)$ models, which is shown in 
Fig. \ref{aniso2}, and now both models are fitted well in both low and high 
energy regimes. The corresponding $\chi^2$ values depict these all. The 
analysis of the anisotropy parameter $\Delta$ as a function of the magnetic 
field strength $B$ for different CR energies in Fig.~\ref{aniso3} highlights 
the influence of both magnetic and gravitational effects on CRs propagation. 
The results demonstrate that MTG models, such as 
$f(R, T) = \alpha R + \beta e^{T}$ and $f(R, T) = R + f_0 RT^{\delta}$, 
predict lower anisotropy compared to the standard $\Lambda$CDM model. These 
findings emphasize the sensitivity of anisotropy to underlying gravitational 
frameworks and various factors like the magnetic field, offering a potential 
avenue to distinguish between standard and MTGs through observations of UHECRs.

\begin{figure}[h!]
\centerline{
\includegraphics[scale=0.5]{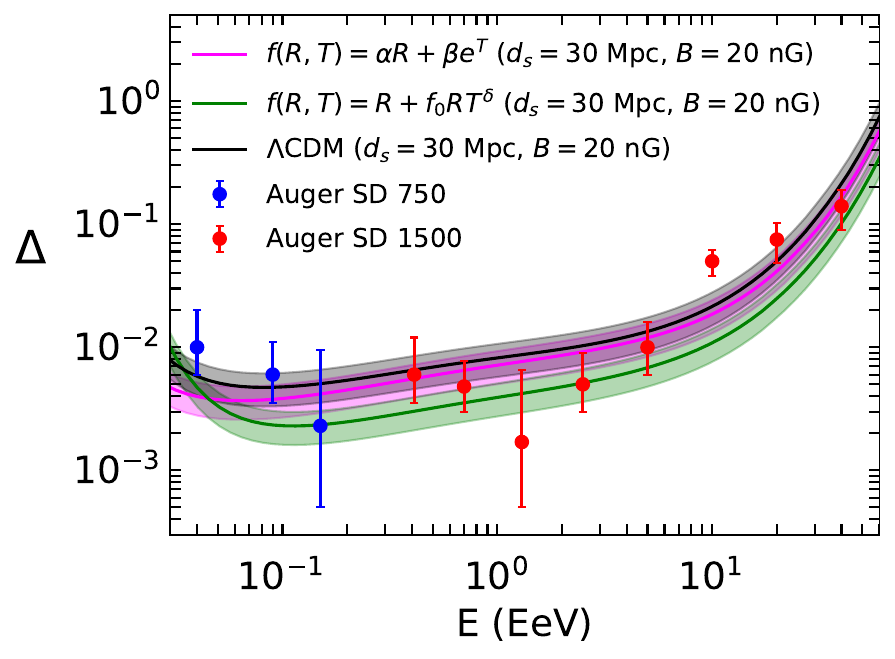}\hspace{0.5cm}
\includegraphics[scale=0.5]{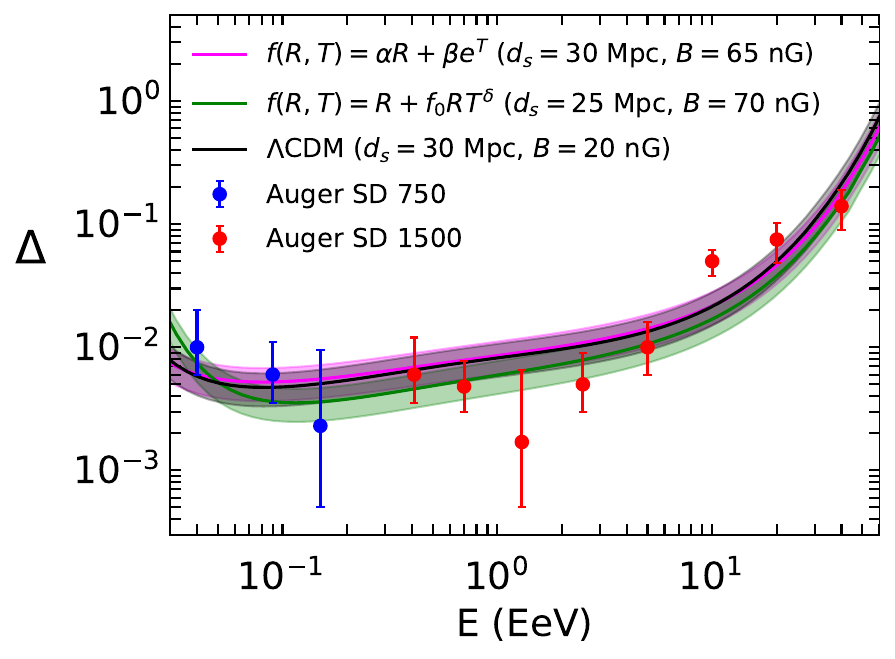}}
\vspace{-0.2cm}
\caption{Left: The unmodified anisotropy of the UHECRs as a function of energy 
$E$ for the considered $f(R, T$) models for $d_\text{s} =30$ Mpc and $B=20$ nG 
with the uncertainty colour bands. Right: The modified anisotropy of the 
UHECRs as a function of energy $E$ for $d_\text{s} =30$ Mpc, $B=65$ nG and 
$d_\text{s} =25$ Mpc, $B=70$ nG for $f(R, T) = \alpha R + \beta e^{T}$  and 
$f(R, T) = R + f_0 R T^{\delta}$ respectively, with the uncertainty colour 
bands. The observational data are taken from the Auger \cite{apj891}.}
\label{aniso4}
\end{figure}

We summarised the results of this study in Fig.~\ref{aniso4}, wherein the left 
panel represents the comparison between the predictions of the $f(R, T)$ 
models and the $\Lambda$CDM model, all calculated for $d_\text{s} = 30$ Mpc and 
$B = 20$ nG. Uncertainty bands are added to both $f(R, T)$ gravity models 
to account for potential variations in model parameters. The experimental data 
from Auger SD 750 and SD 1500 are well accommodated by the predictions, with 
the shaded regions indicating a reasonable overlap. The right panel of 
Fig.~\ref{aniso4} showcases the results of the $f(R, T)$ models with 
optimized parameter sets ($d_\text{s} = 30$ Mpc, $B = 65$ nG for $f(R, T) = 
\alpha R + \beta e^{T}$, and $d_\text{s} = 25$ Mpc, $B = 70$ nG for 
$f(R, T) = R + f_0 R T^{\delta}$), along with the $\Lambda$CDM prediction for 
$d_\text{s} = 30$ Mpc, $B = 20$ nG. The adjustments in $B$ and $d_\text{s}$ 
for the $f(R, T)$ models further enhance their compatibility with the Auger 
datasets. The addition of uncertainty bands provides a robust representation of 
theoretical predictions and demonstrates the sensitivity of anisotropy to 
the theoretical parameters. The inclusion of uncertainty bands in the updated 
predictions emphasizes the sensitivity of anisotropy to the models' 
assumptions. These results underline the importance of precise tuning of 
$f(R, T)$ parameters, as changes in $B$ and $d_\text{s}$ significantly 
influence the 
alignment with data. Moreover, the ability of $f(R, T)$ gravity models to 
suppress anisotropy at intermediate energies compared to $\Lambda$CDM suggests 
that modified gravity effects could play a vital role in explaining CRs 
propagation and their interactions in extragalactic magnetic fields.
Overall, our results indicate that the $f(R, T)$ gravity models can 
effectively reproduce the observed energy-dependent anisotropy of UHECRs 
incorporating the effects of magnetic fields and source separations. 

It is important to note that the results presented here in fact 
constitute a first approximation, as the cosmological parameters used here 
are based on the $\Lambda$CDM estimates. A more rigorous treatment would be 
required to derive the cosmological parameters specifically for the $f(R, T)$ 
models, either through early-universe data such as the CMB or late-universe 
observations. However, since the primary aim of this work is the study of the 
pattern of CR anisotropies rather than detailed cosmological parameter 
estimation, we leave such an in-depth analysis for a future investigation.
\section*{Acknowledgements} UDG is thankful to the Inter-University Centre for 
Astronomy and Astrophysics (IUCAA), Pune, India for the Visiting Associateship 
of the institute.


\end{document}